# Systematic induction of superconductivity in $Y_{1-x}Ca_xBa_2Cu_3O_{6.3}$ system

V.P.S. Awana[*], Anurag Gupta, M.A. Ansari, S.B. Samanta, R.B. Saxena,

and H. Kishan

National Physical Laboratory K.S. Krishnan Marg, New Delhi 110012, India.

and

Devendra Buddhikot and S.K. Malik

Tata Institute of Fundamental Research Homi Bhabha Road, Mumbai 400005, India.

V. Ganesan and A.V. Narlikar

Inter-University Consortium for DAE Facilities, University Campus, Khandwa

Road, Indore-452017, MP, India.

Compounds of series $YBa_2Cu_3O_{7-\delta}$ ($\delta \approx 0.7$), being synthesized by solid-state synthesis route with x = 0.0, 0.10, 0.15 and 0.20, crystallize in single-phase form with tetragonal structure (space group *P*4/*mmm*). The *c*-lattice parameter increases with increasing x, indicating successful substitution of $Y^{3+}$ by bigger $Ca^{2+}$ ion. Resistance versus temperature (R vs. T) measurements show that pristine sample (x = 0.0) is semiconducting down to 5 K. Induction of superconductivity is seen with an increase in x. The x = 0.10 sample exhibit the onset of superconducting transition ($T_c^{onset}$) at around 42 K without attaining zero resistance superconducting transition temperature ($T_c^{R=0}$) state down to 5 K. For x = 0.15 and 0.20 samples $T_c^{R=0}$ is observed at 25 K and 35 K with $T_c^{onset}$ of 52 K and 61 K respectively. Thermo-electric power (S) measurements in temperature range of 5 – 300 K, exhibited $T_c^{S=0}$ at around 25 K and 35 K respectively for x = 0.15 and 0.20 samples. Further a sign change in S from +ve to –ve is observed at low temperatures for x = 0.0 sample. Room temperature +ve S value decreases with increase in x, indicating enhanced number of mobile holes. The Magneto-transport measurements under applied magnetic fields of 3 and 6 Tesla show the broadening of superconducting transition for x = 0.15, and 0.20 samples. For x = 0.10 sample the superconducting transition is suppressed in applied fields of 3 and 6 Tesla. Our results will help in constructing the complete phase diagram of $Y_{1-x}Ca_xBa_2Cu_3O_{7-\delta}$ compounds with various x and $\delta$.

*PACS:* 74.25.-q, 74.25.Dw, 74.25.Fy., 74.62.-c, 74.62.Bf.

Corresponding Author: E-mail: awana@mail.nplindia.ernet.in

**I. INTRODUCTION**

The phase diagram of $RBa_2Cu_3O_{7-\delta}$ (R = rare earth elements, except Ce, Pr and Tb), as a function of oxygen content, $\delta$, shows that for $\delta=1$ these compounds are antiferromagnetic (AFM) insulators with Cu moments ordering above room temperature. With decrease in $\delta$ i.e, on increasing the overall oxygen content of the system, the mobile p-type carriers are injected which bring the system from AFM insulator to a superconducting state with $T_c \sim 90$ K ($\delta \approx 0.0$). In the intermediate region of $1.0 < \delta < 0.0$ with variable value of p-type carriers, the AFM ordering temperature of Cu spins decreases and eventually vanishes at a particular $\delta$ value and superconductivity starts appearing [1,2].

Another way of exploring the phase diagram of $RBa_2Cu_3O_{7-\delta}$ (R-123) system is via partial substitution of $R^{3+}$ by $Ca^{2+}$, within the solubility limit, which is around 20-30% only [3-8]. Substitution of trivalent Y by divalent Ca in $Y_{1-x}Ca_xBa_2Cu_3O_{7-\delta}$ directly dopes mobile carriers in the system, provided $\delta$ does not change with x [8]. The $Y^{3+}$ site $Ca^{2+}$ substitution in $Y_{1-x}Ca_xBa_2Cu_3O_{7-\delta}$ system with $\delta \approx 0.0$, decreases $T_c$ primarily due to over-doping for $x < 0.05$. The same happens for $x > 0.05$ by increasing values of $\delta$ [4,8]. On the other hand when $Ca^{2+}$ is substituted at $Y^{3+}$ site in under-doped ($\delta > 0.0$) $Y_{1-x}Ca_xBa_2Cu_3O_{7-\delta}$ system, it increases the $T_c$ of pristine un-doped system [5,9,10]. For $\delta$ values close to 1.0, one achieves the insulator to metal transition and subsequently superconductivity by this substitution [5,9]. For example superconductivity is reported with $T_c \sim 20$K in $Y_{0.8}Ca_{0.2}Ba_2Cu_3O_{6.1}$ system [5]. Interestingly most of the studies regarding the induction of superconductivity by $Y^{3+}$- site $Ca^{2+}$ substitution in insulating Y-123 are carried out near highly under-doped regime i.e. close to $\delta \approx 1.0$ [4,7,8,11-13]. On the other hand suppression of superconductivity in $Y_{1-x}Ca_xBa_2Cu_3O_{7-\delta}$ system is mainly reported for $\delta \approx 0.0$. The composition $Y_{1-x}Ca_xBa_2Cu_3O_{7-\delta}$ with intermediate range of $\delta$ between 1.0 and 0.0 has not been examined fully [14]. In the present study we explore the $Y_{1-x}Ca_xBa_2Cu_3O_{7-\delta}$ system with $\delta \approx 0.70$ and report its superconducting characteristics. Superconductivity is introduced systematically in $Y_{1-x}Ca_xBa_2Cu_3O_{7-\delta}$ system ($\delta \approx 0.70$) at lower concentration of Ca than those reported earlier for $\delta = 1.0$ systems. Our present results will help in constructing the complete phase diagram of $Y_{1-x}Ca_xBa_2Cu_3O_{7-\delta}$ compounds with various values of x and $\delta$.

## II. EXPERIMENTAL DETAILS

Samples of the series $Y_{1-x}Ca_xBa_2Cu_3O_{7-\delta}$ with x = 0.0, 0.10, 0.15 and 0.20 were synthesized by a solid-state reaction route using ingredients $Y_2O_3$, $CaCO_3$, $BaCO_3$ and CuO. Calcinations were carried out on mixed powders at 880 $^0$C, 890 $^0$C, 900 $^0$C and 910 $^0$C each for 24 hours with intermediate grindings. Pressed pellets were annealed in a flow of oxygen at 920 $^0$C for 40 hours and subsequently cooled slowly to room temperature with an intervening annealing for 24 hours at 600 $^0$C. These pellets were further annealed in flowing $N_2$ gas at 600 $^0$C for 24 hours and subsequently cooled to room temperature. X-ray diffraction (XRD) patterns were obtained at room temperature (MAC Science: MXP18VAHF[22]; Cu$K_\alpha$ radiation). Magnetization measurements were performed on a SQUID magnetometer (Quantum

Design: MPMS-5S). Resistivity measurements under applied magnetic fields of up to 6 T were made in the temperature range of 5 to 300 K using a four-point-probe technique (PPMS, Quantum Design). Thermoelectric power (TEP) measurements were carried out by dc differential technique over a temperature range of 5 – 300 K, using home made set up. Temperature gradient of ~ 1 K is maintained throughout the measurement.

### III. RESULTS AND DISCUSSION

Figure1 depicts the X-ray diffraction patterns for all samples of the series $Y_{1-x}Ca_xBa_2Cu_3O_{7-\delta}$ with x = 0.0, 0.10, 0.15 and 0.20. All these samples crystallize in a tetragonal structure of (space group $P4/mmm$), without any detectable impurities within the x-ray detection limits. Earlier it has been shown that the safe solubility limit of Ca substitution at Y-site in Y-123 is around 20% [11] and our results are consistent with that. The $c$-lattice parameter increases monotonically with increasing x, indicating successful substitution of $Y^{3+}$ by bigger ion $Ca^{2+}$. The $c$-lattice parameter of the pristine x = 0.0 sample is 11.803(3) Å, which facilitates us in knowing that the oxygen content in the system is 6.3 [15]. The ionic size of $Ca^{2+}$ in eight-fold coordination number is 1.12 Å, while that of $Y^{3+}$ in the same co-ordination is 1.02 Å. The system remains tetragonal over the whole range of doping (20% of $Ca^{2+}$ at $Y^{3+}$). Monotonic increase of $c$-lattice parameter with x in $Y_{1-}$ $_x$Ca$_x$Ba$_2$Cu$_3$O$_{7-\delta}$ system guarantees the substitution of Ca$^{2+}$ at Y$^{3+}$ site in the same co-ordination number of eight. The lattice parameters of various samples are given in Table 1. Across the series the *a*-lattice parameter is nearly constant with slight decreasing trend. This is due to the fact that though the *a*-lattice parameter is supposed to increase slightly due to bigger ion Ca substitution, the increasing number of carriers decrease the in plane Cu(2)–O(2) distance and hence the former effect is nullified. It is known that increasing p-type carriers in HTSC compounds increase the hybridization of the in-plane Cu(3d) and O(2p) orbitals resulting in a decrease both in Cu(2)-O(2) distance and the *a*-lattice parameter [14,16]. This situation is different than the one observed earlier for Y$_{1-x}$Ca$_x$Ba$_2$Cu$_3$O$_{7-\delta}$ system with $\delta \approx 0.0$, where Ca presumably prefers a lower co-ordination number [8]. Interestingly the Ca-doped samples though superconducting, are still tetragonal like un-doped non-superconducting and insulating pristine compound. Earlier there are some reports about the phase transformation of tetragonal under-doped R-123 to orthorhombic with induction of superconductivity by Y$^{3+}$ site Ca$^{2+}$ substitution [17,18]. Our results show that, the Y$_{1-x}$Ca$_x$Ba$_2$Cu$_3$O$_{6.3}$ system remains tetragonal within the solubility limit of Ca irrespective of whether the same is superconducting or not.

Resistance versus temperature plots for Y$_{1-x}$Ca$_x$Ba$_2$Cu$_3$O$_{7-\delta}$ with x = 0.0, 0.10, 0.15 and 0.20 are shown in Fig.2. Pristine sample (x = 0.0) is highly semiconducting and its resistance values in Fig.2 are divided by 6. Though, the absolute resistivity values are not plotted in Fig.2, their resistances were measured on similar pieces, indicating that the scaling factor should be nearly the same. Keeping this in mind, one infers from Fig.2, that both room temperature conductivity and normal state (above $T_c^{onset}$) conduction process of $Y_{1-x}Ca_xBa_2Cu_3O_{7-\delta}$ system improves with increase in Ca concentration. The x = 0.0 sample is semiconducting down to 5 K, for which the $R_{5K}/R_{300}$ ratio is nearly 10. Though x = 0.10 sample exhibits $T_c^{onset}$, its normal state behavior is semiconducting with $R_{onset}/R_{300}$ ratio of nearly 2. This implies that the normal state conduction improves significantly after $Ca^{2+}$ substitution at $Y^{3+}$ site. For x = 0.15 sample, the normal state conduction is mostly metallic in nature with a slight upturn above $T_c^{onset}$. Further it shows a $T_c(R=0)$ at 25 K. The x = 0.20, compound exhibits metallic behavior in the entire temperature range between room temperature and $T_c^{onset}$. The x = 0.20 sample shows maximum $T_c(R=0)$ value of around 35 K. It is clear from our resistance data shown in Fig.2, that superconductivity is induced systematically by $Ca^{2+}$ substitution at the $Y^{3+}$ site in $Y_{1-x}Ca_xBa_2Cu_3O_{7-\delta}$ ($\delta \approx 0.7$) system.

The results of thermoelectric power (S) measurements on $Y_{1-x}Ca_xBa_2Cu_3O_{6.30}$ with x = 0.0, 0.10, 0.15 and 0.20 are shown in Fig.3. The value of S at room temperature (290 K) is found to be positive for all the samples, indicating them to be predominantly hole (p) type conductors. Also the value of $S_{290\ K}$ decreases monotonically with x (Table 1). Implying that the number of mobile p-type carriers increase with increase in x for $Y_{1-x}Ca_xBa_2Cu_3O_{6.3}$ system. For strongly correlated systems the absolute value of S is known to be inversely proportional to the number of mobile carriers [19]. Further with decreasing temperature S passes through a maximum ($S_{max}$) and later starts decreasing with further decrease in temperature. The temperature corresponding to $S_{max}$ i.e. T ($S_{max}$), is marked as a cross in Figure 3. T ($S_{max}$) decreases monotonically with increasing x. Thermoelectric power measurements below T ($S_{max}$), exhibits rather sharp transitions to $T_c^{S=0}$ at around 25 K and 35 K respectively for x = 0.15 and 0.20 samples. A change in sign of S from +ve to –ve is observed at low temperatures for x = 0.0 sample due to which thermoelectric power passes through S = 0 state, which should not be confused with superconductivity. Pristine (x = 0.0) sample is insulating and the same exhibits seemingly a change in carrier sign at low temperatures. For x = 0.10 sample though $T_c^{S=0}$ is observed at around 7 K, clear superconductivity like transition is not seen. This may possibly be indicative of the weak superconductivity for x = 0.10 sample. In brief one can conclude that thermoelectric power measurements corroborate the resistance versus temperature results shown in Fig.2.

Magneto transport in applied fields of 0, 3 and 6 Tesla for x = 0.10 sample is shown in Fig.4. No apparent MR is seen in normal state i.e. above $T_c^{onset}$.

This indirectly excludes the possibility of any magnetic ordering of Cu-spins above $T_c^{onset}$. $T_c^{onset}$ is not observed in applied fields of 3 and 6 Tesla, indicating towards weaker superconductivity. Interestingly the S vs T measurements did not exhibit clear superconducting transition for this sample, see Fig.3. Hence both R vs T and S vs T measurements indicate towards weak superconductivity in x = 0.10 sample. In optimally doped HTSC compounds, the $T_c^{onset}$ generally remains unchanged under magnetic field. Sharp decrease of $T_c^{onset}$ under applied magnetic field indicates towards the presence of non-uniform distribution of mobile carriers in various grains of the system. This type of systems may eventually form the SNS (superconducting-normal-superconducting) junctions, origin of which may be related to the intra-grain phase-lock transitions, which are susceptible to decrease in $T_c^{onset}$ under magnetic field [20].

Fig.5 depicts the magneto-transport measurements for x = 0.15 sample. This sample exhibits $T_c$(R=0) at around 25 K without any applied magnetic field. Under applied magnetic field of 3 and 6 Tesla, R = 0 state is not observed and $T_c^{onset}$ is found to decrease with field. Decreasing $T_c^{onset}$ under magnetic field calls for the similar explanation (SNS/SIS junctions formation due to non-uniform distribution of holes with in different grains) as for x = 0.10 sample. It means that even in x = 0.15 sample having $T_c$(R=0) of around 25 K, the distribution of mobile carriers is not uniform. No apparent MR is seen in normal state i.e. above $T_c^{onset}$, which excludes the possibility of any magnetic ordering of Cu-spins above $T_c^{onset}$ in this sample.

The magneto-transport measurements for x = 0.20 sample are shown in Fig.6. This sample shown $T_c(R=0)$ of around 35 K. Considering that fact that for higher x values disorder plays a negative role towards superconductivity, the $T_c$ of this sample indeed should be higher [8,21]. Both doped carriers and the disorder due to substitution seems to compete with each other. This is the reason that one never achieves the optimum $T_c$ of the system, even with sufficient substitution level. Under applied magnetic fields of 3 and 6 Tesla, the $T_c(R=0)$ decreases to around 10 K and 8 K. All the $T_c$ (superconducting transition) values are given in Table 1. $T_c^{onset}$ of this sample remains nearly unchanged under applied magnetic fields of 3 and 6 Tesla. This situation is similar to that as observed for other HTSC compounds. Interestingly though $T_c^{onset}$ is unchanged under magnetic field, a step is seen in broadened transition, which is again reminiscent of some SNS junctions in the compound. No apparent MR is seen in normal state excluding the possibility of any magnetic ordering of Cu-spins above $T_c^{onset}$. We would like to mention that though a small step like structure is seen in broadened transition of x = 0.20 sample, its unchanged $T_c^{onset}$ under magnetic field warrants more uniform distribution of mobile carriers in various grains than other Ca doped samples. In fact the small step in broadened transition may be due to the presence of natural weak-links in the polycrystalline compound.

## SUMMARY

We have synthesized $Y_{1-x}Ca_xBa_2Cu_3O_{7-\delta}$ ($\delta \approx 0.7$) with x = 0.0, 0.10, 0.15 and 0.20 and induced systematically superconductivity in pristine semiconducting compound. Magneto-transport measurements exhibit unusual broadening of superconducting transition under magnetic field, indicating towards non-uniform distribution of mobile carriers in these systems. Our present results will help in constructing the complete phase diagram of $Y_{1-x}Ca_xBa_2Cu_3O_{7-\delta}$ system as a function of x and $\delta$.

**FIGURE CAPTIONS**

Figure 1. X-ray diffraction patterns of various $Y_{1-x}Ca_xBa_2Cu_3O_{7-\delta}$ samples.

Figure 2. Resistance (R) vs. Temperature (T) for various $Y_{1-x}Ca_xBa_2Cu_3O_{7-\delta}$ samples in temperature range of 5 –300 K.

Figure 3. Thermoelectric power (S) vs. Temperature (T) for various $Y_{1-x}Ca_xBa_2Cu_3O_{7-\delta}$ samples in temperature range of 5 –300 K.

Figure 4. Resistance (R) vs. Temperature (T) in 0, 3, and 6 T fields for $Y_{0.90}Ca_{0.10}Ba_2Cu_3O_{6.3}$ in temperature range of 5 –300 K.

Figure 5. Resistance (R) vs. Temperature (T) in 0, 3, and 6 T fields for $Y_{0.85}Ca_{0.15}Ba_2Cu_3O_{6.3}$ in temperature range of 5 –300 K.

Figure 6. Resistance (R) vs. Temperature (T) in 0, 3, and 6 T fields for $Y_{0.80}Ca_{0.20}Ba_2Cu_3O_{6.3}$ in temperature range of 5 –300 K.

Table 1. Lattice parameters *a*, and *c* and superconducting temperature, $T_c(R=0)$ and superconducting onset temperature $T_c^{onset}$ (K) in applied field of 0, 3 and 6 Tesla for $Y_{1-x}Ca_xBa_2Cu_3O_{6.3}$.

| x | *a* (Å) | *c* (Å) | S(290 K) μV/K | $T_c(R=0)/T_c^{onset}$ H = 0 Tesla | $T_c(R=0)/T_c^{onset}$ H = 3 Tesla | $T_c(R=0)/T_c^{onset}$ H = 6 Tesla |
|---|---|---|---|---|---|---|
| x = 0.0 | 3.859(5) | 11.803(3) | 126 | -/- | -/- | -/- |
| x = 0.10 | 3.854(4) | 11.814(4) | 93 | -/42 K | -/- | -/- |
| x = 0.15 | 3.851(2) | 11.819(2) | 52 | 25/52 K | -/30 | -/24 |
| x = 0.20 | 3.848(5) | 11.823(2) | 46 | 35/61 K | 10/56 K | 8/52 K |

# REFERENCES

1. J.M. Tranquada, D.E. Cox, W. Kunnmann, H. Moudden, G. Shrine, M. Suenaga, P. Zolliker, D. Vaknin, S.K. Sinha, M.S. Alvarez, A.J. Jacobson, and D.C. Jonhston, Phys. Rev. Lett. 60, 156 (1988).

2. J. Rossat-Mignod, P. Burlet, M.J.G.M. Jurgens, J.Y. Henry, and C. Vettier, Physica C 152, 19 (1988).

3. A. Manthiram, S.-J. Lee, and J.B. Goodenough, J. Solid State Chemistry 73, 278 (1988).

4. Y. Tokura, J.B. Torrance, T.C. Huang, and A.I. Nazzal, Phys. Rev. B 38, 7156 (1988).

5. E.M. McCarron III, M.K. Crawford, and J.B. Parise, J. Solid State Chemistry 78, 192 (1989).

6. C. Gledel, J.-F. Marucco, and B. Touzelin, Physica C 165, 437 (1990).

7. B. Fisher, I. Genossar, C.G. Kuper, L. Patlagan, G.M. Reisner, and A. Knizhnik, Phys. Rev. B 47, 6054 (1993).

8. V.P.S. Awana, and A.V. Narlikar, Phys. Rev. B 49, 6353 (1994).

9. P. Starowicz, J. Sokolowski, M. Balanda, and A. Szutula, Physica C 363, 80 (2001).

10. J.W. Radcliffe, N. Athanassopoulou, J.M. Wade, J.R. Cooper, J.L. Tallon, and J.W. Loram, Physica C 235-240, 1415 (1994).

11. K. Hatada, and H. Shimizu, Physica C 304, 89 (1998).

12. T. Watanabe, M. Fujiwara, and N. Suzuki, Physica C 252, 100 (1995).

13. K. Widder, D. Berner, J. Munzel, H.P. Geserich, M. Klaser, G. Muller-Vogt, and Th. Wolf, Physica C 267, 254 (1996).

14. V.P.S. Awana, S.K. Malik, and W.B. Yelon, Physica C 262, 272 (1996).

15. R.J. Cava, A.W. Hewat, E.A. Hewat, B. Batlogg, M. Marezio, K.M. Rabe, J.J. Krajewski, W.F. Peck, and L.W. Rupp, Physica C 165, 419 (1990).

16. M.H. Whangbo and C.C. Torardi, Science 249, 1143 (1990).

17. K.M. Pansuria, U.S. Joshi, D.G. Kuberkar, G.J. Baldha, and R.G. Kulkarni, Solid State Communications 98, 1095 (1996).

18. Amish G. Joshi, D.G. Kuberkar, and R.G. Kulkarni, Physica C 320, 87 (1999).

19. J.R. Cooper, B. Alavi, L.W. Zhou, W.P. Beyermann, and G. Gruner, Phys. Rev. B 35, 8794 (1987).

20. B. Lorenz, Y.Y. Xue, R.L. Meng, and C.W. Chu, Phys. Rev. B 65, 17405 (2002)

21. S.R. Ghorbani, M. Andersson, and O. Rapp, Physica C 390, 160 (2003).

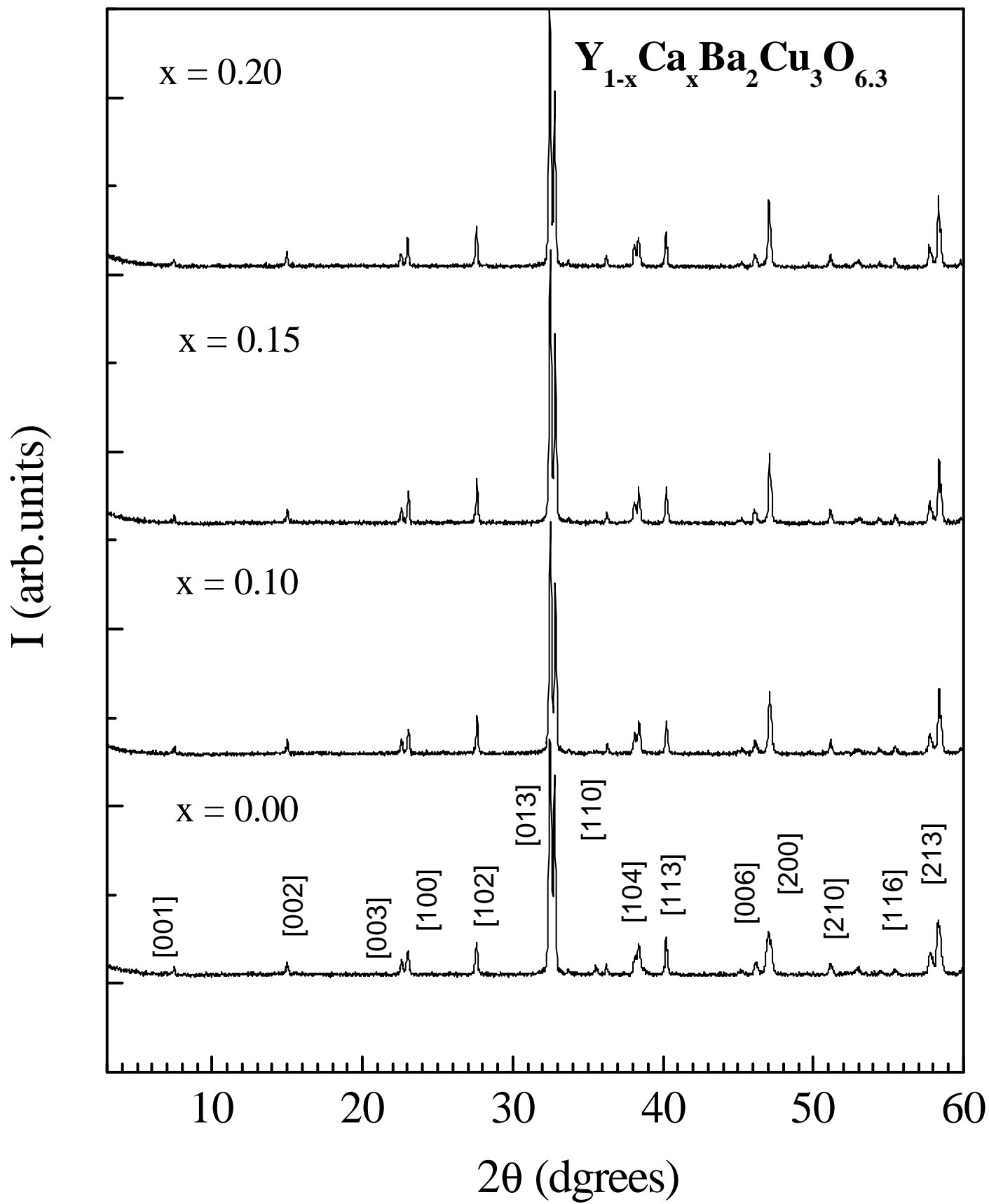

Fig.2 (Awana etal.)

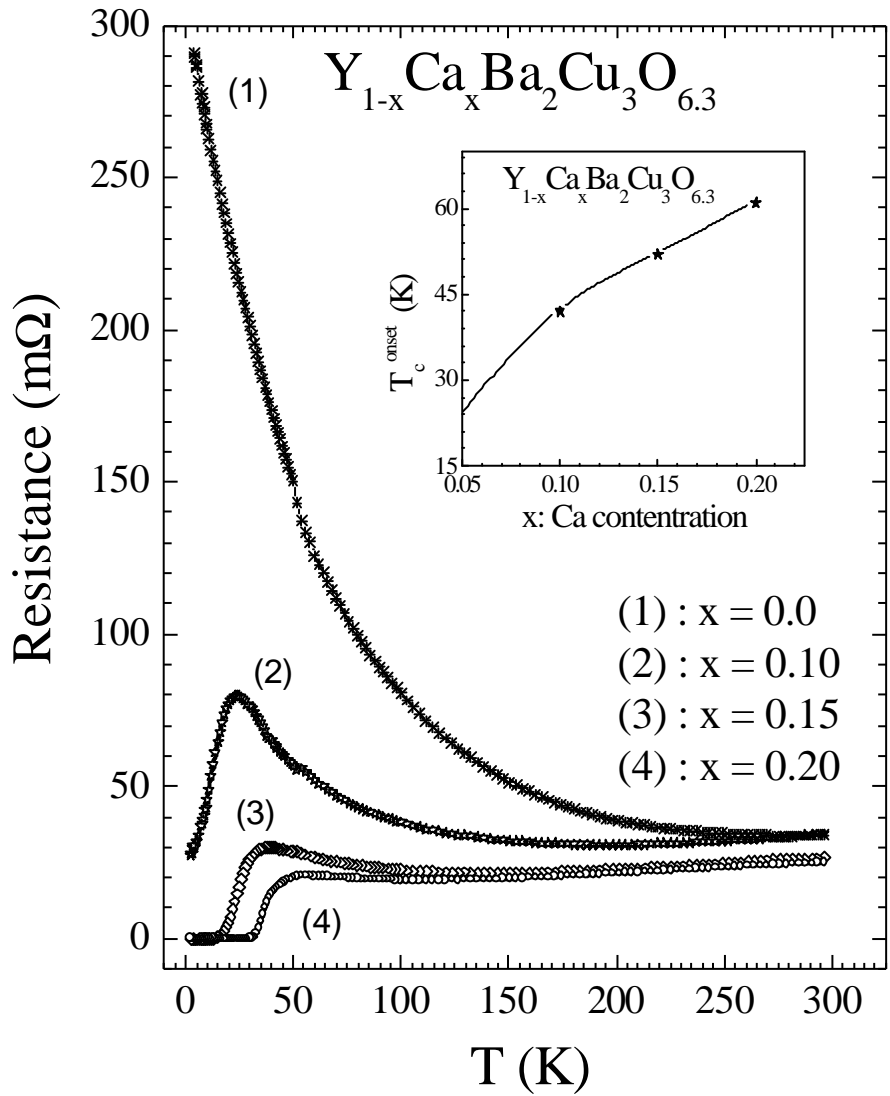

Fig.3 (Awana etal.)

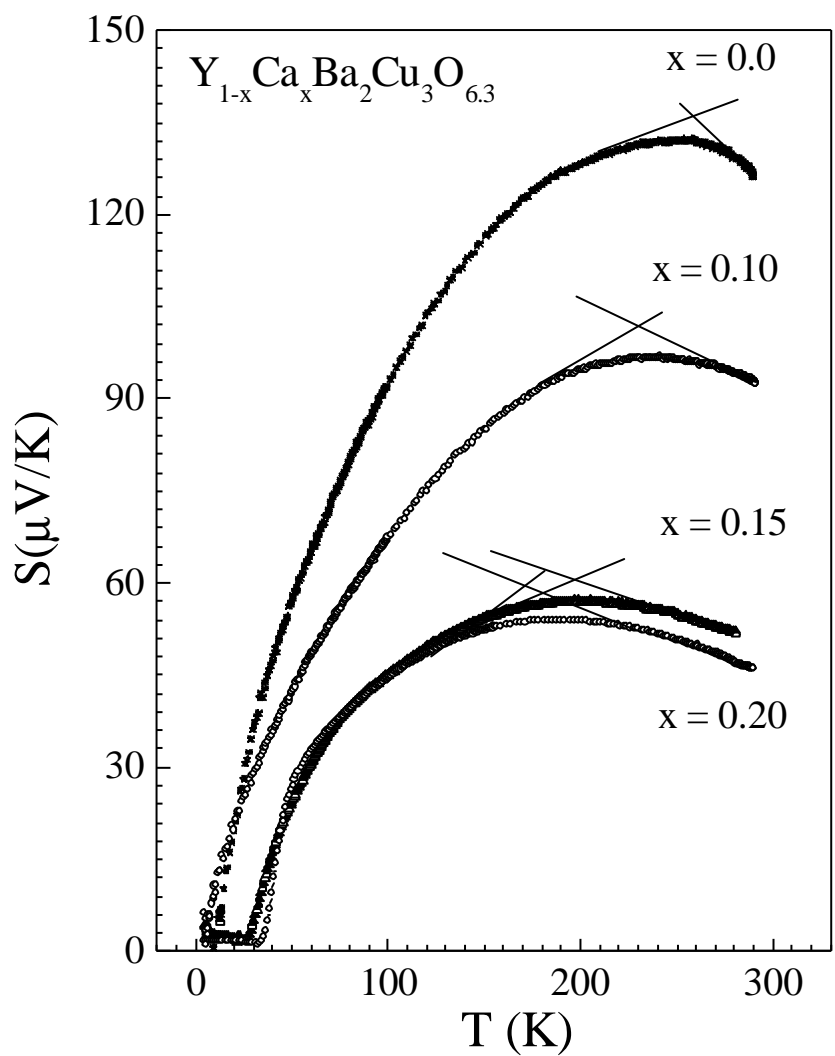

Fig.4 (Awana etal.)

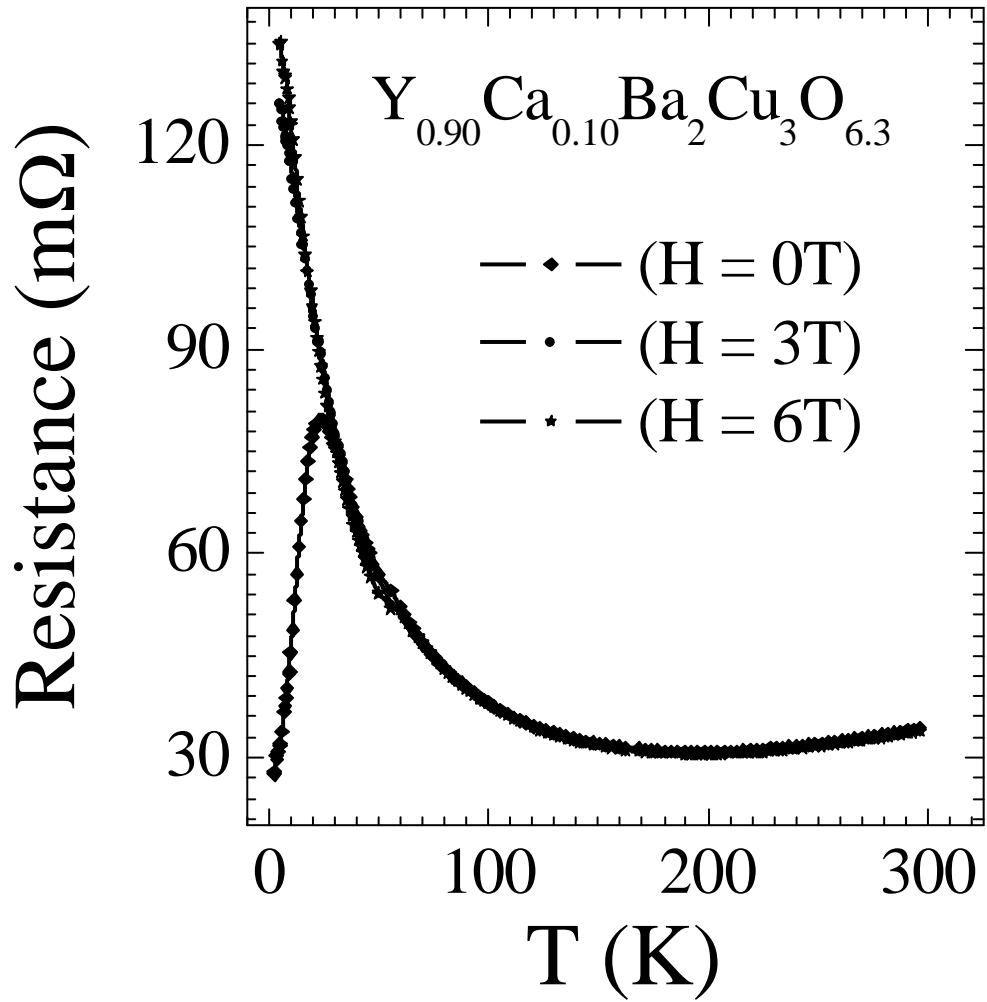

Fig.5 (Awana etal.)

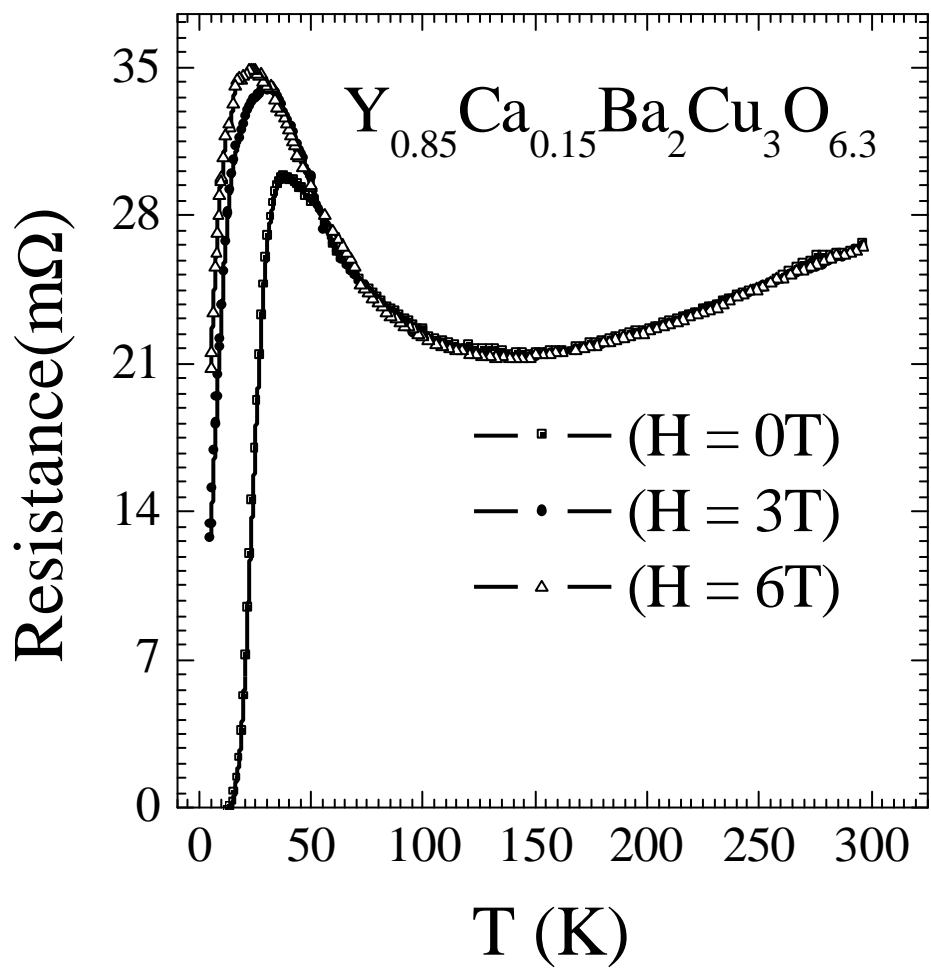

Fig.6 (Awana etal.)

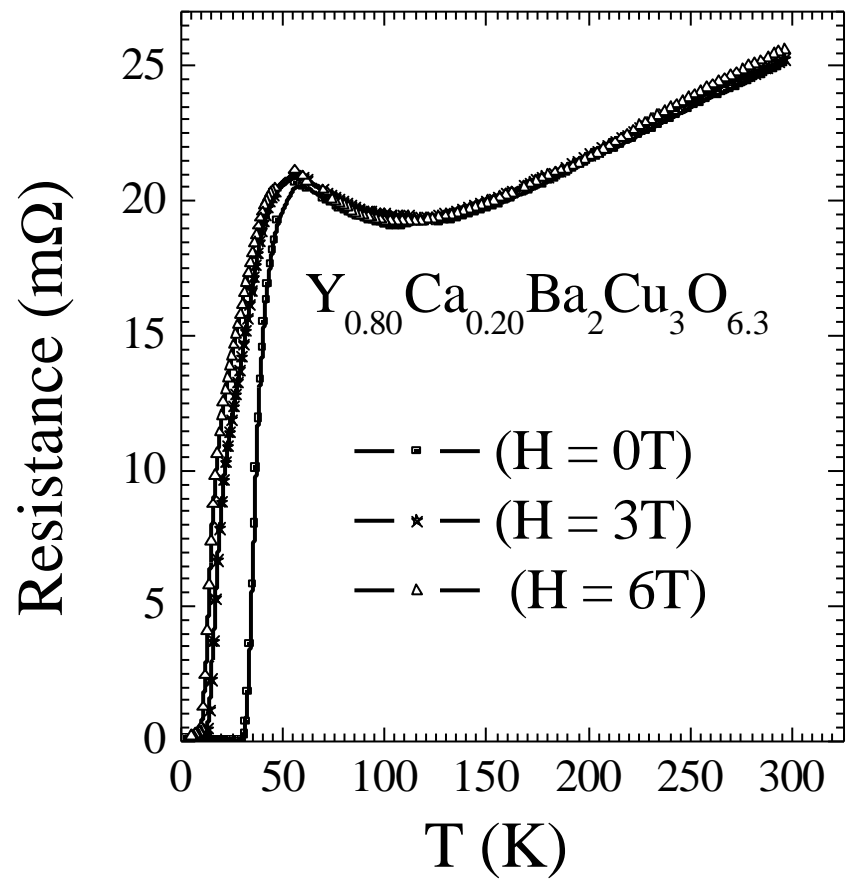